\documentclass[journal=jpccck,manuscript=article]{achemso}

\usepackage[version=3]{mhchem} 
\usepackage{setspace}

\author{Richard Schier}
\affiliation{Humboldt-Universit\"at zu Berlin, Physics Department and IRIS Adlershof, 12489 Berlin, Germany}
\author{Ana M. Valencia}
\affiliation{Humboldt-Universit\"at zu Berlin, Physics Department and IRIS Adlershof, 12489 Berlin, Germany}
\altaffiliation{Carl von Ossietzky Universit\"at Oldenburg, Institute of Physics, 26129 Oldenburg, Germany}
\author{Caterina Cocchi}
\affiliation{Humboldt-Universit\"at zu Berlin, Physics Department and IRIS Adlershof, 12489 Berlin, Germany}
\altaffiliation{Carl von Ossietzky Universit\"at Oldenburg, Institute of Physics, 26129 Oldenburg, Germany}
\email{caterina.cocchi@uni-oldenburg.de}

\title{Microscopic Insight into the Electronic Structure of BCF-Doped Oligothiophenes from \textit{Ab initio} Many-Body Theory}

\begin{document}

\begin{tocentry}
\includegraphics[width=8.25 cm]{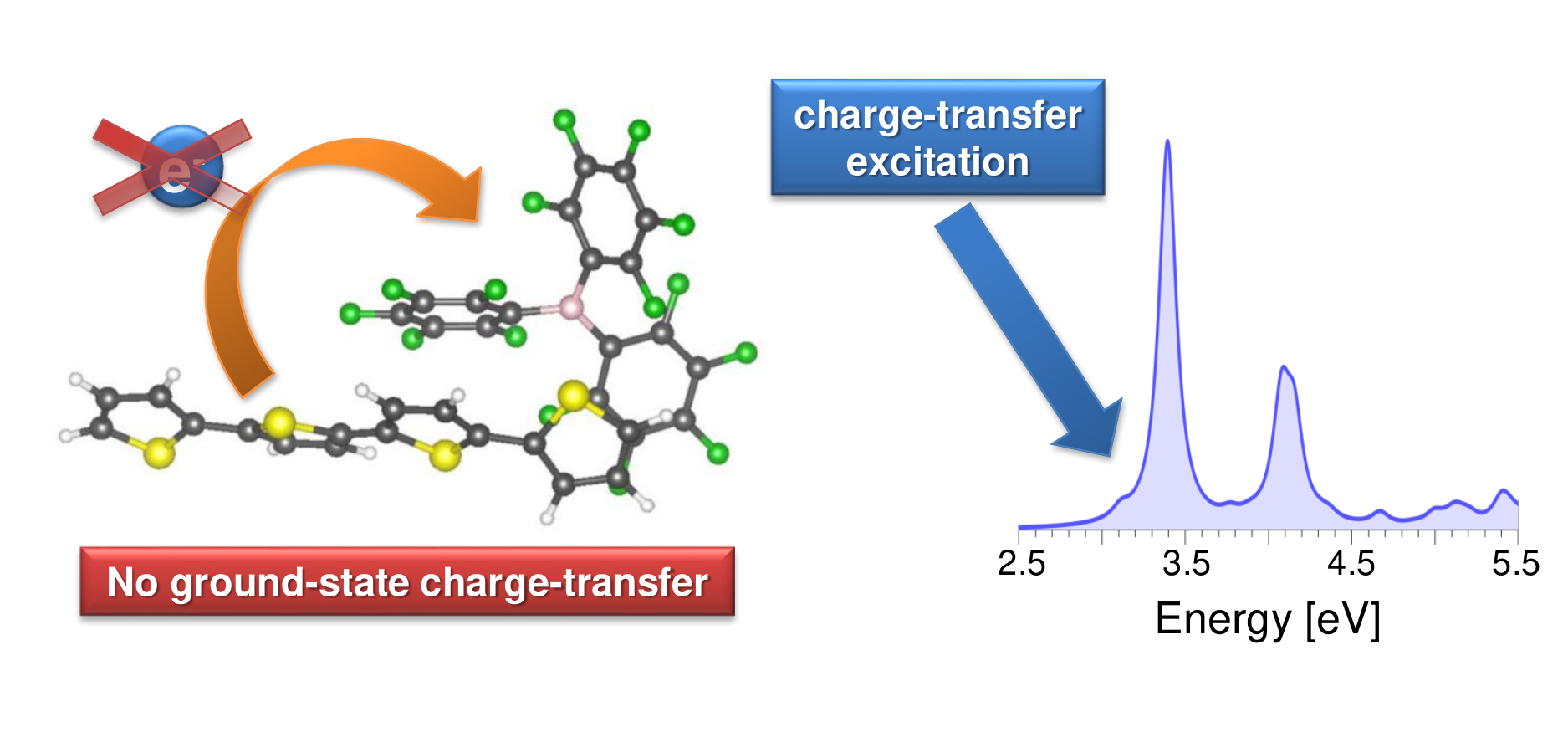}




\end{tocentry}

\begin{abstract}
Lewis acids like tris(pentafluorophenyl)borane (BCF) offer promising routes for efficient $p$-doping of organic semiconductors. The intriguing experimental results achieved so far call for a deeper understanding of the underlying doping mechanisms. In a first-principles work, based on state-of-the-art density-functional theory and many-body perturbation theory, we investigate the electronic and optical properties of donor/acceptor complexes formed by quarterthiophene (4T) doped by BCF. For reference, hexafluorobenzene (\ce{C6F6}) and \ce{BF3} are also investigated as dopants for 4T. Modelling the adducts as bimolecules \textit{in vacuo}, we find negligible charge transfer in the ground state and frontier orbitals either segregated on opposite sides of the interface (4T:BCF) or localized on the donor (4T:\ce{BF3}, 4T:\ce{C6F6}). In the optical spectrum of 4T:BCF, a charge-transfer excitation appears at lowest-energy, corresponding to the transition between the frontier states, which exhibit very small but non-vanishing wave-function overlap. In the other two adducts, the absorption is given by a superposition of the features of the constituents. Our results clarify that the intrinsic electronic interactions between donor and acceptor are not responsible for the doping mechanisms induced by BCF and related Lewis acids. Extrinsic factors, such as solvent-solute interactions, intermolecular couplings, and thermodynamic effects, have to be systematically analyzed for this purpose.
\end{abstract}
\newpage
\section{Introduction}

Doping represents a crucial process to enable the application of organic semiconductors in optoelectronics~\cite{ostr16cr,jaco-moul17am,wang+18csr}. 
A consolidated consensus acknowledges the formation of charge-transfer complexes~\cite{zhu+11cm,salz+13prl,kim+16jpcc,tiet+18natcom,beye+19cm,vale-cocc19jpcc,vale+20pccp} and ion-pairs~\cite{brau-sala07cpl,moul+08cm,yim+08am,ping+10jpcl,ping-nehe13prb,bao+14afm,mend+15ncom} as the dominant doping mechanisms in organic semiconductors~\cite{mend+15ncom,salz-heim15jesrp,salz+16acr}. 
The emergence of Lewis acids like tris(pentafluorophenyl)borane, \ce{B(C6F5)3} (in short BCF)~\cite{mass-park64joc,kort+17acie} as novel dopant species has opened new routes for efficient $p$-doping of organic polymers and oligomers~\cite{welc+09jacs,welc-baza11jacs,zala+12acie,pove+14jacs,zala+14am,ping+16aelm,hans+16acie,lin+16mrc,yan+18aem,pani+18as,yura+19cm,phan+19jpcl,yura+19natm,mans+20jmcc}. 
The pioneering study by Pingel \textit{et al.}~\cite{ping+16aelm}, has identified integer charge transfer between poly(3-hexylthiophen-2,5-diyl) (P3HT) and BCF, but has not been able to fully disclose the underlying physical processes. 
The recent work by Yurash \textit{et al.}~\cite{yura+19natm} has demonstrated that the doping mechanisms induced by BCF and other Lewis acids is largely mediated by the protonation of a portion of the polymer in solution. 
Mansour and coworkers~\cite{mans+20jmcc} have just shown that BCF promotes in P3HT the formation of polarons with different characteristics depending on the morphology and aggregation of the donor species. 

The complexity unraveled by these observations calls for an in-depth understanding of the quantum-mechanical interactions that characterize these complexes. 
This is a necessary step to gain insight into intrinsic properties of the materials, such as orbital hybridization, wave-function overlap, and, more generally, electronic interactions, which strongly affect the behavior of the system in the actual experimental and device conditions, but are not straightforward to be detected in the measurements. 
In order to disclose and rationalize these effects, we present in this paper a state-of-the-art first-principles investigation of the electronic and optical properties of the donor/acceptor complex formed by a quaterthiophene (4T) oligomer doped by BCF. 
In our analysis, performed in the framework of hybrid density-functional theory and many-body perturbation theory (\textit{GW} approximation and Bethe-Salpeter equation), we investigate isolated adducts \textit{in vacuo}. 
In this way, we are able to pinpoint the electronic interactions within the complex and to understand the underlying quantum-mechanical effects. 
For comparison, we investigate two additional complexes formed by the Lewis acid \ce{BF3} and by hexafluorobenzene (\ce{C6F6}). 
In the ground state, we focus on the charge transfer, the level alignment between the donor and acceptor, and the character of the frontier orbitals in the adducts. 
Furthermore, we compute the optical absorption spectra of the complexes and quantitatively determine the spatial distribution of the electron and hole densities. With this analysis we are able to provide relevant information about the intrinsic characteristics of these prototypical donor/acceptor complexes, which is necessary for an improved understanding of the fundamental doping mechanisms induced by BCF and related Lewis acids.

\section{Methodology}
\subsection{Theoretical Background}
The results presented in this work are carried out from first principles in the framework of density functional theory (DFT)~\cite{hohe-kohn64pr} and many-body perturbation theory (MBPT)~\cite{onid+02rmp}, including the $GW$ approximation and the solution of the Bethe-Salpeter equation (BSE). In DFT, the many-body system is mapped into the fictitious Kohn-Sham (KS) system of non-interacting electrons~\cite{kohn-sham65pr}. The electronic states are described by the wavefunctions $\phi_j$, which are the solutions of the secular equation with effective Hamiltonian $\hat{h}$:
\begin{equation}
\hat{h}\phi_j = (\hat{T}+\hat{V}_{eff})\phi_j = (\hat{T}+\hat{V}_{ext}+\hat{V}_{H}+\hat{V}_{xc})\phi_j =     \epsilon_j \phi_j
\end{equation}
The eigenvalues $\epsilon_j$ correspond to the energies of the respective $\phi_j$. $\hat{T}$ is the kinetic energy operator and $V_{eff}$ the effective KS potential.
This term is given by the sum of the external potential $V_{ext}$, corresponding to the interaction between the electrons and the nuclei, the Hartree potential $V_{H}$, and the exchange-correlation potential $V_{xc}$, describing the exchange-correlation interaction between the electrons.
We recall that the exact form of the last term is unknown and, therefore, has to be approximated.

The DFT results represent the starting point for the calculation of the quasi-particle (QP) electronic structure within the $GW$ approximation~\cite{hedi65pr}, where the electronic self-energy is given by $\Sigma = iGW$. Here, we use the perturbative $G_0W_0$ approach~\cite{hybe-loui85prl} to solve the QP equation
\begin{equation}
    \epsilon_j^{QP}=\epsilon_j^{KS}+\langle\phi_j^{KS}|\Sigma(\epsilon_j^{QP})-\hat{V}_{xc}|\phi_j^{KS}\rangle
\end{equation}
and obtain the QP energies $\epsilon_j^{QP}$.
Finally, to compute the optical excitations, we solve the BSE, which is the equation of motion of the two-particle polarizability~\cite{stri88rnc}:
\begin{equation}
L = L_0 +  L_0 \Xi L,
\end{equation}
where $L$ is the interacting electron-hole correlation function related to the two-particle Green's function, $L_0$ is its non-interacting counterpart, and $\Xi$ is the electron-hole interaction kernel including the statically screened Coulomb interaction as well as the exchange potential between the positively-charged hole and the negatively-charged electron.
In practice, the problem is mapped into a secular equation with an effective two-particle Hamiltonian including the BSE kernel~\cite{brun+16cpc}.

\subsection{Computational Details}

Equilibrium geometries are calculated by force minimization in the framework of DFT. For these calculations we use the all-electron code FHI-aims~\cite{blum+09cpc}. The Perdew-Burke-Ernzerhof semi-local functional~\cite{perd+96prl} is used to approximate the exchange-correlation potential, together with tight integration grids and TIER2 basis sets~\cite{havu+09jcp}. The Tkatchenko-Scheffler scheme~\cite{tkat-sche09prl} is adopted to include van der Waals interactions. The optimization procedure is carried out until the Hellmann-Feynman forces are smaller than 10$^{-3}$~eV/\AA. 

To calculate the electronic and optical properties in the framework of DFT and MBPT, we use the MOLGW code~\cite{brun+16cpc}. Gaussian-type cc-pVDZ~\cite{brun12jcp} basis sets are used and the resolution-of-identity approximation~\cite{weig+02} is applied. Here, the hybrid exchange-correlation functional using the Coulomb-attenuating method CAM-B3LYP~\cite{yana+04cpl} is chosen. The Bader charge analysis scheme is used to compute the partial charges~\cite{bade90oup,henk+06cms,sanv+07jcc,tang+09jpcm}.
The $G_0W_0$ and the BSE calculations are performed including all the occupied states and with approximately three time as many unoccupied states. The total number of occupied and unoccupied states is determined by the number of basis functions used to calculate the KS wave-functions. In the case of 4T:BCF, this amounts to a total of 822 KS states, including 209 occupied and 613 virtual orbitals. For 4T:\ce{C6F6} 130 occupied and 384 virtual orbitals are used, while in 4T:\ce{BF3}, 101 occupied and 301 virtual ones are adopted.
The BSE is solved in the Tamm-Dancoff approximation. 
The spatial distribution of the $\lambda$th electron-hole pair is evaluated from the electron and hole densities, defined as:
\begin{equation}
    \rho_{h}^{\lambda}(\mathbf{r})=\sum_{\alpha\beta}A_{\alpha\beta}^{\lambda}|\phi_{\alpha}(\mathbf{r})|^2
\end{equation}
and
\begin{equation}
    \rho_{e}^{\lambda}(\mathbf{r})=\sum_{\alpha\beta}A_{\alpha\beta}^{\lambda}|\phi_{\beta}(\mathbf{r})|^2  \text{  ,}
\end{equation}
respectively~\cite{cocc+11jpcl,deco+14jpcc,vale-cocc19jpcc,vale+20pccp}, where
$\phi_{\alpha}$ and $\phi_{\beta}$ are the occupied and the unoccupied QP states contributing to the $\lambda$th excitation, respectively. The weighting coefficients $A_{\alpha\beta}^{\lambda}$ are the absolute squares of the normalized BSE eigenvectors.

\section{Results}

\subsection{Structural properties}

\begin{figure}
    \centering
  \includegraphics[width=0.5\textwidth]{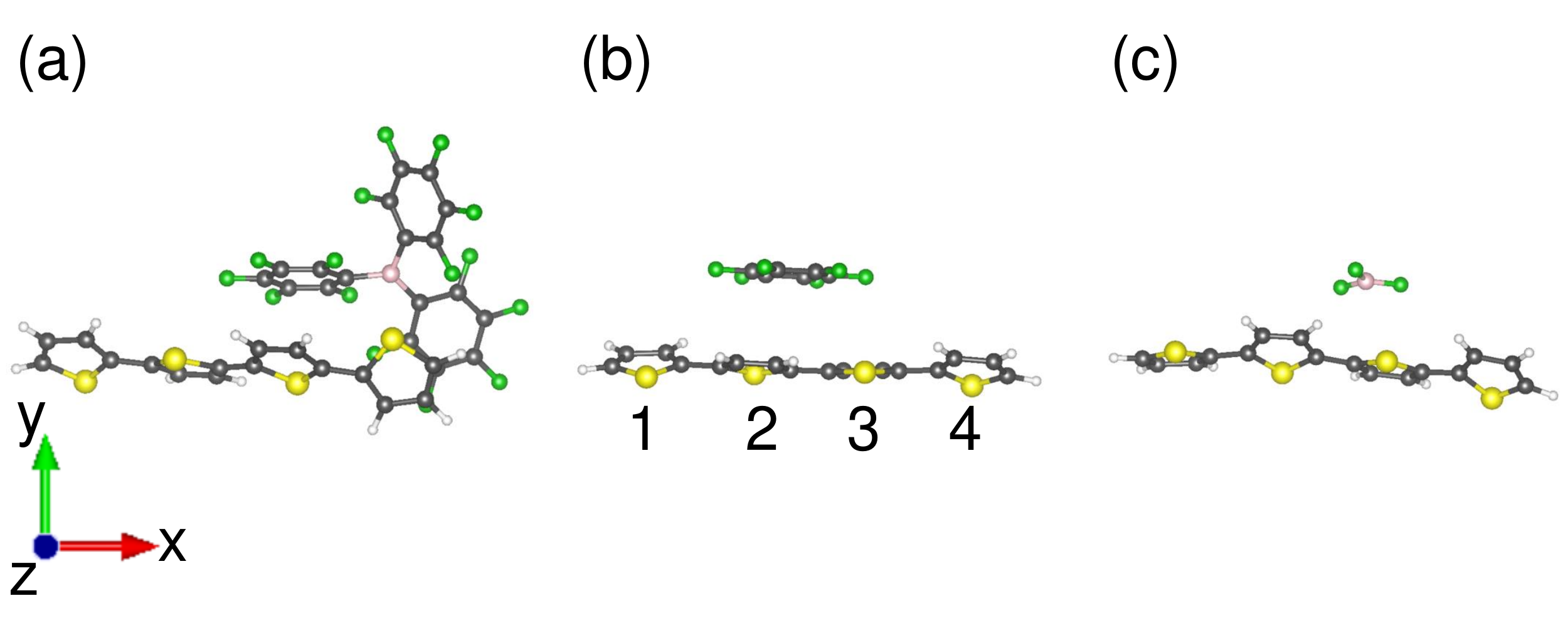}
	\caption{Ball-and-stick representation of the adducts considered in this work: (a) 4T:BCF, (b) 4T:C$_6$F$_6$ and (c) 4T:BF$_3$ complexes. Sulphur, carbon, hydrogen, fluorine, and boron atoms are represented in yellow, grey, white, green, and pink, respectively. The number of the 4T rings in panel (b) is used in Table~\ref{tab:angle}.}
	\label{fig:struct}
\end{figure}

To study the impact of BCF on the electronic and optical properties of organic semiconductors, we consider a model system formed by 4T interacting with one BCF molecule in the gas phase, in absence of any reference of such a system in the solid state.
It is worth mentioning, however, that in co-crystals~\cite{sato+19jmcc} the geometries of faced 4T and tetracyanoquinodimethane is not significantly affected compared to the charge-transfer complex in the gas phase~\cite{mend+15ncom,vale-cocc19jpcc,vale+20pccp}.
4T is a representative thiophene oligomer absorbing visible light~\cite{lap+97jpca,fabi+05jpca,sieg+11pccp,cocc-drax15prb}, which is often used to mimic short segments of polythiophene chains~\cite{mend+15ncom}.
In the relaxed geometry obtained for 4T:BCF and reported in Figure~\ref{fig:struct}a), the acceptor is adsorbed on one side of the donor. This geometry resembles the adduct formation proposed by Pingel \textit{et al.} \cite{ping+16aelm}, with the boron atom aligned to the sulphur on one of the outer rings of 4T.
An alternative geometry with BCF on the center of the 4T backbone was also explored but turned out to be energetically less favorable than the one in Figure~\ref{fig:struct}a).

For reference, we also investigate the adducts formed by 4T doped by the Lewis acid \ce{BF3}, as well as by hexafluorobenzene (\ce{C6F6}).
The optimized geometries of the energetically most favorable configurations are shown in Figure~\ref{fig:struct}b)-c). 
Additional information about the structural properties of the adducts is given in Table~\ref{tab:angle}, where the values of the dihedral angle S-C-C-S are reported for each ring of 4T, according to the labeling in Figure~\ref{fig:struct}b).
We recall that in the isolated 4T molecule each angle is equal to 180$^{\circ}$.
It is evident that the interaction with BCF causes a significant distortion in the 4T backbone such that the thiophene ring directly below the acceptor (ring 4), is subject to a 40$^{\circ}$ torsion with respect to the adjacent ring 3 (see Table~\ref{tab:angle}), which is facing the \ce{C6F5} unit adsorbed directly above it.
The rest of the 4T molecule also experiences a distortion ranging from 25$^{\circ}$ to 13$^{\circ}$ further away from the dopant adsorption site. 
Also the adsorption of \ce{BF3} causes a pronounced distortion of the thiophene backbone with a torsion angle of approximately 20$^{\circ}$ between the rings.
On the other hand, when interacting with \ce{C6F6}, the 4T backbone remains almost planar except for a twist of about 25$^{\circ}$ of the side rings (see Table~\ref{tab:angle}). 
In all adducts, the bond lengths undergo negligible changes compared to isolated 4T, which are therefore not reported herein.

\begin{table}
\caption{Dihedral angle between the different 4T rings (as labeled in Figure \ref{fig:struct}).}
\label{tab:angle}
\begin{tabular}{c|ccc}
4T-rings & 4T:BCF   & 4T:C$_6$F$_6$ & 4T:BF$_3$ \\ \hline
1	-	2	&	156.22$^{\circ}$	&	151.99$^{\circ}$	&	156.60$^{\circ}$		\\
2	-	3	&	167.18$^{\circ}$	&	173.81$^{\circ}$	&	159.01$^{\circ}$	\\
3	-	4	&	140.39$^{\circ}$	&	156.14$^{\circ}$	&	156.66$^{\circ}$	\end{tabular}
\end{table}

The mutual distance between donor and acceptor varies in the three adducts, due to the different nature of the dopant molecule involved. 
In the case of 4T:BCF, the shortest B-S distance amounts to 3.75~\AA{} while in 4T:\ce{BF3} it is equal to 3.83~\AA{}.
The F atoms in the latter system are separated by 3.42~\AA{} from the underlying C atoms in 4T.
In the 4T:\ce{C6F6} complex, the carbon and fluorine atoms of the acceptor lie on the same plane, about 3.4~\AA{} above the 4T backbone.
This distance is comparable with the separation between the \ce{C6F5} ring and the donor in 4T:BCF.


\subsection{Electronic properties}
\begin{table}
\caption{Bader charges on the acceptors in the considered adducts. The mean values of the charges on the \ce{C6F5} rings in BCF and on the F atoms in \ce{BF3} are reported. Charges are given in units of electrons with positive (negative) values indicating change depletion (accumulation).}
\label{tab:bader_da}
\begin{tabular}{cccc}
      Species                         & \multicolumn{1}{c}{4T:BCF} & \multicolumn{1}{c}{4T:\ce{C6F6}} & \multicolumn{1}{c}{4T:\ce{BF3}} \\ \hline \hline
Tot. charge & -0.068 & -0.027 & -0.032 \\
\hline
B  & +2.955  & - & +2.957   \\
\ce{C6F5} / F  & -1.008  & - & -0.996 \\
\hline
\end{tabular}
\end{table}

As the next step in our analysis, we examine the electronic properties of the adducts, starting from inspecting the charge transfer in the ground state. 
For this purpose, we make use of the Bader charge analysis, partitioning the system between the donor and the acceptor units (see Table~\ref{tab:bader_da}).
In all three adducts we find a negligible charge transfer between 4T and the respective acceptor species, of the order of 10$^{-2}$ electrons.
In the case of the 4T:BCF and 4T:\ce{BF3}, a significant charge transfer occurs \textit{within} the acceptor molecules.
This behavior is not driven by the interacting 4T molecule but is an intrinsic property of the Lewis acids \ce{BF3} and BCF, as shown in more details in the Supporting Information (Tables~S1 and S2), where the partial charges of the single acceptor molecules are reported.
The B atom, coordinated with strong electron-withdrawing species, such as F atoms and \ce{C6F5} rings, donates almost in full its three valence electrons (see Table~\ref{tab:bader_da}). 
The absence of ground-state charge transfer in 4T:BCF, modeled here as an isolated bimolecule \textit{in vacuo}, is in perfect agreement with a previous calculation performed in an idealized thiophene polymer doped by BCF~\cite{mans+20jmcc}. 
This finding reveals that the doping mechanism between 4T and BCF does not result from direct charge transfer, but is driven by external factors, such as polaron formation, integer charge transfer with consequent protonation of the donor, and solution environment~\cite{ping+16aelm,gaul+18natm,yura+19natm,mans+20jmcc} that are not included in these calculations. 

\begin{figure*}
	\centering
  \includegraphics[width=0.9\textwidth]{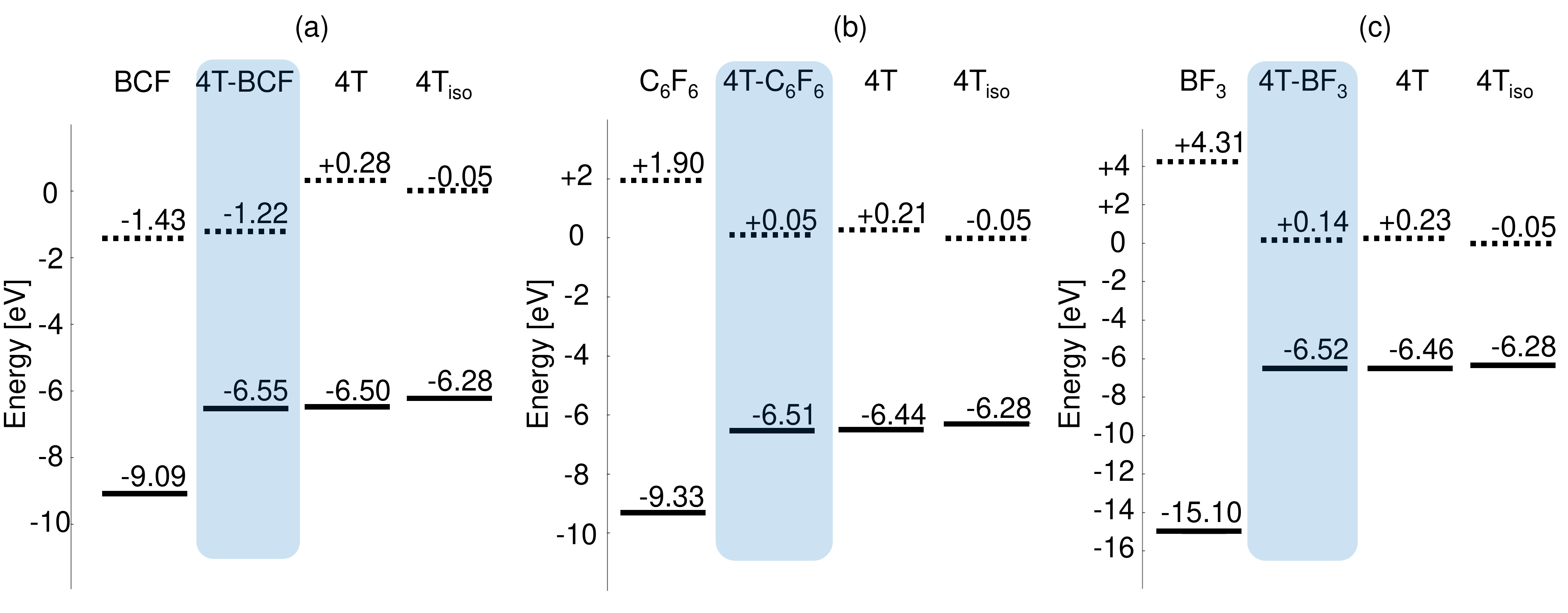}
	\caption{Level alignment, computed from $G_0W_0$, of (a) 4T:BCF (b) 4T:\ce{C6F6} (c) 4T:\ce{BF3}, with respect to their constituents: The frontier states of the acceptors are referred to the gas-phase molecule, while for 4T we report both the frontier energies in the isolated molecule (4T$_{\text{iso}}$) and in the distorted geometry of the adduct.}
		\label{fig:lev}
\end{figure*}

\begin{figure}
	\centering
  \includegraphics[width=0.5\textwidth]{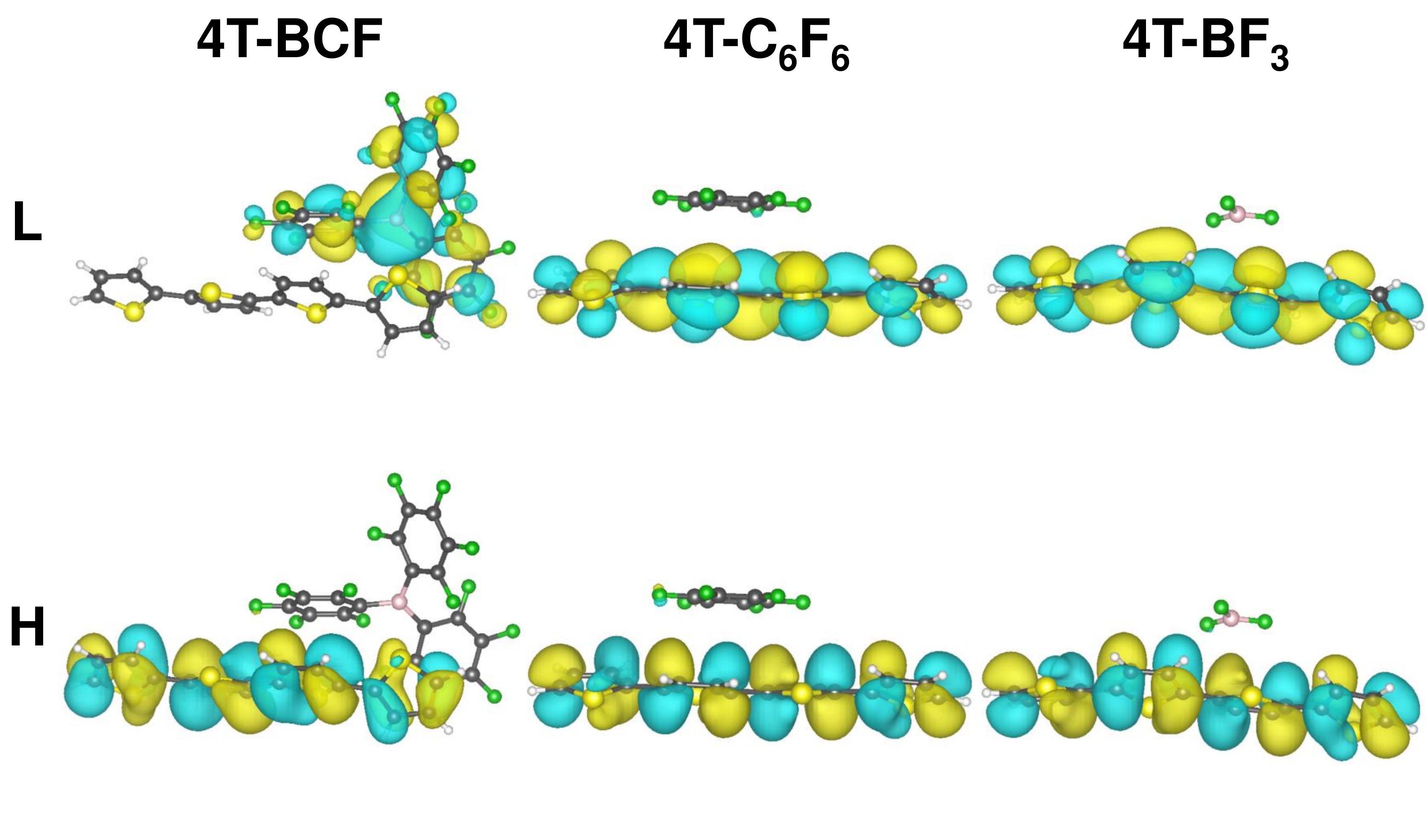}
	\caption{HOMO (H) and LUMO (L) of the considered D/A complexes with isosurfaces depicted at 10\% of their maximum value.}
		\label{fig:MOs_DA}
\end{figure}

We continue with our analysis on the electronic properties of the adducts by considering the energy level alignment computed from $G_0W_0$ on top of hybrid DFT, as reported in Figure~\ref{fig:lev}.
In each panel, from left to right, we show the frontier states of the isolated acceptor, of the adduct, of the donor in the relaxed geometry of the respective adduct, as well as in the gas phase (4T$_{\text{iso}}$).
In the case of 4T:BCF, the lowest unoccupied molecular orbital (LUMO) of BCF is energetically comprised between the frontier states of the donor, resulting in a \textit{type II} level alignment.
A close inspection of Figure~\ref{fig:lev}a) reveals that the energies of the highest-occupied molecular orbital (HOMO) of 4T:BCF is only a few tens meV lower than the HOMO of the donor in the geometry of the adduct. 
This difference increases up to a few hundreds meV with respect to the isolated 4T (4T$_{\text{iso}}$).
The LUMO of 4T:BCF is equally close in energy to the LUMO of BCF alone.
These correspondences are mirrored by the character of the frontier orbitals in the adduct, shown in Figure~\ref{fig:MOs_DA}, left panel.
The HOMO is localized on the donor, while the LUMO on the acceptor.
Both states retain the character of the corresponding orbitals of the constituents (see Figure~S1 in the Supporting Information).
A careful analysis of the LUMO of 4T:BCF reveals, however, a slight hybridization between donor and acceptor, which will play a role in the optical properties discussed below.

4T:\ce{C6F6} and 4T:\ce{BF3} are instead characterized by a \textit{type I} level alignment between the constituents (see Figure~\ref{fig:lev}b,c). 
In those cases, the relatively small size of the acceptor molecules gives rise to band gaps of the order of 11~eV for \ce{C6H6} and almost 20 eV for \ce{BF3}, which largely exceed the one of 4T. 
As a result, the frontier orbitals of the adducts coincide with those of the donor, both in terms of energy (see Figure~\ref{fig:lev}) and of spatial distribution (see Figure~\ref{fig:MOs_DA}b,c).
Also the QP gap is almost identical to the one of 4T ($\sim$6.5~eV) which is more than 1~eV larger compared to the one of 4T:BCF (5.3~eV). 
In the 4T:\ce{C6F6} and 4T:\ce{BF3} differences in the QP gaps of the adducts compared to those of the donor are ascribed to polarization effects exerted by the acceptor molecule. The calculated larger gap is in agreement with previous findings for breaking the planarity of the $\pi$-conjugated 4T molecule~\cite{gierschner07}.

A final note about the effect of the distortion of 4T in the presence of the dopant.
Contrary to previous results obtained for edge-functionalized graphene nanoflakes~\cite{cocc+11jpcc,cocc+12jpcc}, where backbone distortions give rise to a band-gap reduction, in the three adducts considered here, the HOMO and the LUMO of the donor in the geometry of the complex are energetically (slightly) lower and higher, respectively, compared to those of their flat and isolated counterparts. 
This behavior can be interpreted as an indication that the planar geometry predicted by semi-local DFT for gas-phase 4T is not the global minimum of this configuration, as also suggested by earlier experimental results~\cite{dice+99jpca,kouk+00jcp}.


\subsection{Optical properties}

\begin{figure}[h!]
    \centering
\includegraphics[width=0.48\textwidth]{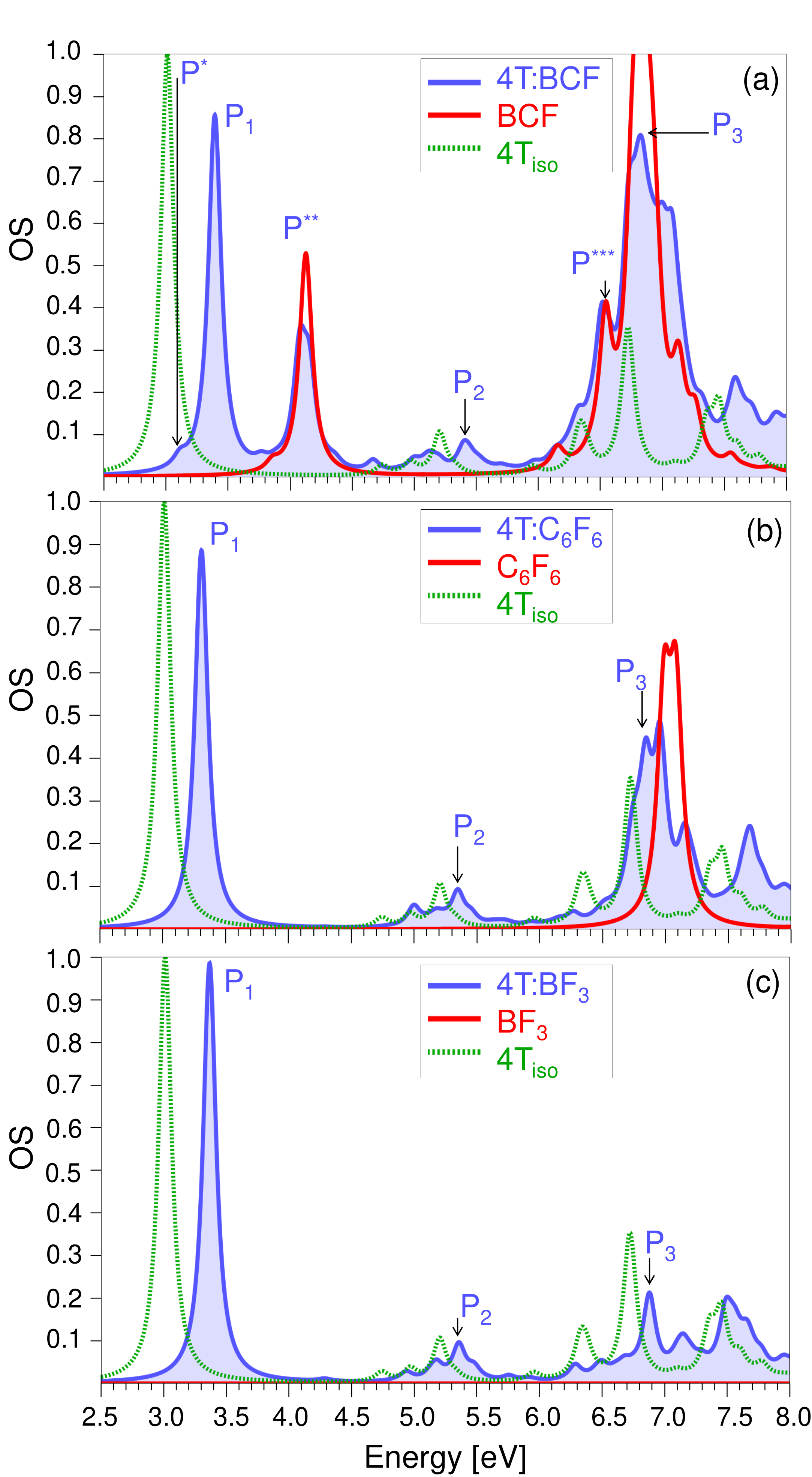}
\caption{Optical spectra of (a) 4T:BCF, (b) 4T:\ce{C6F6}, (c) 4T:\ce{BF3} and of their respective constituents in the equilibrium geometries as isolated compounds. A Lorentzian broadening with a full width at half maximum of 0.125 eV is applied to the spectra to mimic the excitation lifetime. The strength of all peaks is normalized to the height of first maximum in the spectrum of 4T.}
    \label{fig:spectra}
\end{figure}

The analysis of the electronic structure reported above provides all the ingredients to investigate the optical properties of the adducts. 
In Figure~\ref{fig:spectra} we report the absorption spectra computed from the solution of the BSE.
In each panel we show the result for the adducts together with the spectra of their isolated constituents in the respective equilibrium geometries.

\begin{figure*}
    \centering
\includegraphics[width=\textwidth]{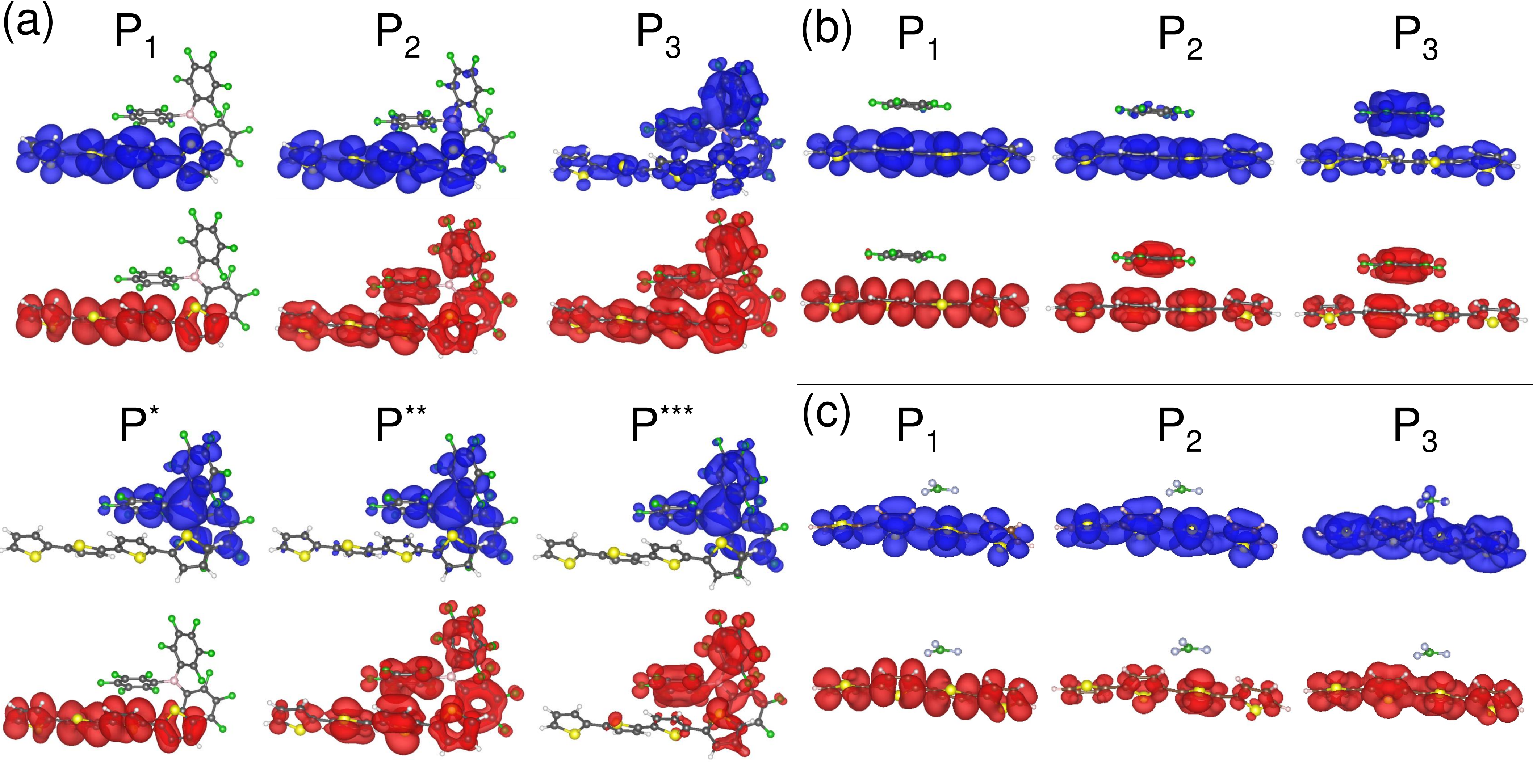}
\caption{Electron (blue) and hole (red) densities of the optical excitations listed in Table 1 for (a) 4T:BCF, (b) 4T:\ce{C6F6}, and (c) 4T:\ce{BF3}. The isosurfaces are plotted at approximately 1\% of their maximum value.}
    \label{fig:eh_DA}
\end{figure*}

The spectrum of 4T:BCF (Figure~\ref{fig:spectra}a) is dominated at the onset by an intense peak at 3.4~eV (P$_1$), including a weak shoulder at 3.1~eV (P*).
Other maxima are found at approximately 4~eV (P**) and at 5.5~eV (P$_2$).
Deeper in the UV region between 6~eV and 7.5~eV a broad and intense feature comprises the peaks P*** and P$_3$.
Comparison between the spectrum of the adduct and the spectra of its individual constituents offers clear indications about the nature of the aforementioned absorption peaks. 
First of all we notice that P$_1$ is blue-shifted by approximately 0.5~eV with respect to the first excitation of 4T, considered in its flat geometry (further details about this spectrum are in the Supporting Information, see Figure~S6). 
This trend is consistent with the electronic structure discussed above, where the QP gap of distorted 4T is larger than the one of its flat counterpart (see Figure~\ref{fig:lev}a).
The 4T-like character of P$_1$ is confirmed by the spatial distribution of the electron and hole densities associated to this excitation (see Figure~\ref{fig:eh_DA}a).
On the other hand, the lowest-energy peak, P*, is a charge-transfer excitation emerging in the spectrum of 4T:BCF due to the character of its frontier orbitals (see Figure~\ref{fig:MOs_DA}). 
In fact, P* corresponds to the transition from the HOMO of the adduct, localized on the donor, to the LUMO, mainly distributed on the acceptor and only slightly hybridized with 4T. 
This minimal wave-function overlap explains the weak but non-vanishing oscillator strength (OS) of P*, in spite of its almost pure charge-transfer character~\cite{cocc+11jpcl}.
The analysis of the electron and hole densities in Figure~\ref{fig:eh_DA}a) supports this interpretation. 

The second bright peak in the spectrum of 4T:BCF, P**, appears at the same energy as the first maximum in the spectrum of BCF. 
The analysis of the corresponding electron and hole densities (Figure~\ref{fig:eh_DA}a) reveals, however, that this excitation in the adduct is rather delocalized across the whole complex, especially as far as the hole is concerned. 
P$_2$ appears in proximity of the second absorption maximum in the spectrum of 4T (for further details, see Supporting Information, Figure~S6 and related discussion). 
However, the analysis of the corresponding electron and hole densities (see Figure~\ref{fig:eh_DA}), which are largely delocalized across the donor and the acceptor molecules, reveals that this excitation is evidently a new feature emerging in the adduct. 
Finally, P*** and P$_3$ appear in the UV region, where also an intense and equally broad feature is present in the spectrum of BCF alone. 
P*** is mainly localized on the acceptor and the corresponding electron and hole densities are resembling of their counterparts in the excitation of isolated BCF at the same energy (see Supporting Information, Figure~S8). 
On the other hand, P$_3$ is a delocalized excitation across the whole adduct, as clearly visible through the plot of the corresponding electron and hole densities in Figure~\ref{fig:eh_DA}a).
Its large OS is therefore a direct consequence of such a large wave-function overlap. 

We now turn to the analysis of the spectra of 4T:\ce{C6F6} and 4T:\ce{BF3}, shown in Figure~\ref{fig:spectra}b) and c), respectively. 
At a glance, it is evident that they both resemble the spectrum of 4T:BCF, especially in the low-energy region. 
The first intense peak, P$_1$, is found at approximately the same energy in the spectra of all three adducts and bears the same 4T-like character regardless of the dopant species, as shown by the corresponding electron and hole density distributions in Figure~\ref{fig:eh_DA}.
Also P$_2$ appears in all three spectra at about the same energy (5.5~eV) and with approximately the same OS. 
However, the character of this excitation varies depending on the acceptor. 
In the case of 4T:\ce{C6F6}, the hole is distributed on both 4T and \ce{C6F6} while the electron is localized only on the donor (see Figure~\ref{fig:eh_DA}b).
Conversely, in 4T:\ce{BF3}, both the 
electron and hole densities are localized solely on 4T. 

The higher-energy excitation P$_3$ is found between 6.5~eV and 7.0~eV in the spectra all adducts.
However, like P$_2$, its relative OS and its character are affected by the dopant species. 
In 4T:\ce{C6F6}, P$_3$ is rather intense (its OS is about half of the one of P$_1$), and its electron and hole densities are both distributed across the entire adduct, similar to its counterpart in the spectrum of 4T:BCF (see Figure~\ref{fig:eh_DA}). 
Note that the first peak in the spectrum of isolated \ce{C6F6} is energetically very close to P$_3$ (see Figure~\ref{fig:spectra}b). 
On the contrary, in the spectrum of 4T:\ce{BF3}, P$_3$ is almost as weak as P$_2$ and is again almost entirely distributed on the donor only (see Figure~\ref{fig:eh_DA}c).
A careful inspection of Figure~\ref{fig:spectra}c) reveals that the spectrum of isolated \ce{BF3} does not feature any absorption feature in the region between 2.5~eV and 8.0~eV, considered in this analysis. 
The absorption onset of this molecule is found at 12.23~eV, in line with the existing literature~\cite{dufl+14jpca}, as shown in Figure~S7 of the Supporting Information.

\section{Discussion and Conclusions}
The results presented in this work offer important insight into the intrinsic electronic structure of BCF-doped oligothiophenes, calculated as isolated bimolecules \textit{in vacuo}.
This situation is representative of the local interactions between donor and acceptor species~\cite{mend+15ncom,vale-cocc19jpcc,vale+20pccp}.
However, it is important to consider the fact that in real samples these systems are embedded in an environment (either in solution or in solid-state) and are therefore subject to non-negligible dielectric screening. 
In general, going from the gas- to the condensed-phase induces a broadening and a red-shift of the absorption features as an effect of the dielectric screening exerted by the closely-packed molecules~\cite{humm+04prl,sai+08prb,shar+12prb,cocc-drax15prb,cocc-drax15prb1,cocc+16jcp,rang+16prb,cocc-drax17jpcm}.
Related phenomena like excitonic delocalization are also observed in organic crystals~\cite{schu+07prl,cuda+12prb,cuda+15jpcm}.
In the case of $p$-doped organic crystals in the low-doping limit, it was shown that the screening plays a role in accurately determining ionization potential and electron affinity.~\cite{li+17prm,li+19mh}
It is therefore reasonable to expect a red-shift of a few hundred meV in the computed absorption lines upon inclusion of a screening term mimicking a realistic dielectric embedding for BCF-doped 4T.

In the modeled adducts, we find negligible charge transfer in the ground state of the order of 10$^{-2}$ electrons.
A substantial charge redistribution occurs instead \textit{within} BCF and \ce{BF3}, where the \ce{C6F5} rings and the F atoms, respectively, withdraw almost in full the three valence electrons of the B atom. 
This result supports the recent experimental findings that other mechanisms such as protonation in solution~\cite{yura+19natm} and/or polaron formation~\cite{mans+20jmcc} are responsible for the doping mechanism induced by BCF and related Lewis acids. 

The electronic structure of 4T:BCF is characterized by the \textit{type II} level alignment between its constituents, with the LUMO of BCF comprised between the frontier orbitals of 4T. 
However, different from conventional charge-transfer complexes formed, for example, by 2,3,5,6-tetrafluoro-tetracyanoquinodimethane and 4T (see, \textit{e.g.}, Refs.~\citenum{vale-cocc19jpcc,vale+20pccp}) the HOMO and the LUMO of 4T:BCF are localized on the donor and the acceptor, respectively, with a non-vanishing wave-function overlap between them. 
This behavior is the key for understanding the doping mechanism of 4T:BCF.
In the adducts 4T:\ce{C6F6} and 4T:\ce{BF3}, a \textit{type I} level alignment between the constituents is obtained, owing to the much larger band gap of the acceptor compared to the donor.
Consequently, the frontier orbitals of these adducts coincide with those of 4T alone. 
The optical spectra evidently reflect these characteristics.
In 4T:BCF, the lowest-energy excitation has weak OS and pronounced charge-transfer character, and is energetically very close to the intense peak generated by the optical transition within the donor.
On the other hand, the spectra of the other two adducts result in a mere superposition of the features of their building blocks, consistent with the obtained level alignment and electronic structure.

In conclusion, our analysis reveals that the intrinsic electronic interactions between the donor and the acceptor are not the driving doping mechanism induced by Lewis acids like BCF. 
This main result of our work, although negatively formulated, provides very relevant information about the nature of such donor/acceptor complexes in order to rationalize experimental observations. 
It clarifies, for example, that extrinsic factors, such as solvent-solute interactions, intermolecular coupling among the donors, as well as thermodynamics effects play a crucial role in determining the doping efficiency in these adducts, depending on the specific characteristics of the species involved. 
Further theoretical investigations are certainly needed to address all the aforementioned effects in details.
Our results offer a valuable starting point for this purpose.

\begin{acknowledgement}
Fruitful discussions with Michele Guerrini, Ahmed Mansour, Andreas Opitz, and Dieter Neher are kindly acknowledged.
This work was funded by the German Research Foundation (DFG) through the project ``FoMEDOS'' -- Project number 286798544 (HE 5866/2-1).
Computational resources partly provided by the The North-German Supercomputing Alliance (HLRN) -- project bep00076.
\end{acknowledgement}

\begin{suppinfo}
Additional details about the electronic and optical properties of the adducts and their constituents are provided.

\end{suppinfo}


\providecommand{\latin}[1]{#1}
\providecommand*\mcitethebibliography{\thebibliography}
\csname @ifundefined\endcsname{endmcitethebibliography}
  {\let\endmcitethebibliography\endthebibliography}{}

\end{document}